# Opportunities and Applications of GenAI in Smart Cities: A User-Centric Survey

Dipen Pradhan
dipenp@google.com

Lakshit Arora
lakshit@google.com

Ankit Shetgaonkar
ankiit@google.com

Sanjay Surendranath Girija
sanjaysg@google.com

Shashank Kapoor
shashankkapoor@google.com

Aman Raj
amanraj@google.com

Google

***Abstract*— The proliferation of IoT in cities, combined with Digital Twins, creates a rich data foundation for Smart Cities aimed at improving urban life and operations. Generative AI (GenAI) significantly enhances this potential, moving beyond traditional AI analytics and predictions by processing multimodal content and generating novel outputs like text and simulations. Using specialized or foundational models, GenAI's natural language abilities such as Natural Language Understanding (NLU) and Natural Language Generation (NLG) can power tailored applications and unified interfaces, dramatically lowering barriers for users interacting with complex smart city systems. In this paper, we focus on GenAI applications based on conversational interfaces within the context of three critical user archetypes in a Smart City - Citizens, Operators and Planners. We identify and review GenAI models and techniques that have been proposed or deployed for various urban subsystems in the contexts of these user archetypes. We also consider how GenAI can be built on the existing data foundation of official city records, IoT data streams and Urban Digital Twins. We believe this work represents the first comprehensive summarization of GenAI techniques for Smart Cities from the lens of the critical users in a Smart City.**

## I. Introduction

Nearly 57% of the global population is estimated to be living in urban areas in 2022 [6], and by 2050, the urban population forecasted to grow to 68% by the UN [7]. Managing such large urban areas involves a variety of challenges, some of which can be solved through effective adoption of technology that spans multiple urban subsystems. The concept of Smart City is gaining prominence as a vital strategy for addressing the complexities of urbanization with a goal to improve quality of life for the citizens while fostering sustainable living practices [2]. A Smart City is conceptualized as a complex, integrated system where physical infrastructure, environmental processes, and human activities are seamlessly connected with digital technologies for data-driven insights, management and planning of urban resources and services [1].

Cities are experiencing a convergence of multiple supporting pillars of technology - the proliferation of low-cost internet-enabled devices, ubiquity of internet access in urban spaces, driving the growth of IoT in cities [3]. The general availability of multi-modal GenAI models capable of performing tasks involving multiple disparate data sources [5] now allows the potential creation of new applications based on conversational interfaces for data analysis, exploration and simulation [5]. Natural language interfaces powered by GenAI can offer significant advantages over traditional purpose-built Graphical User Interfaces, which are often limited to specific tasks or datasets. AI adoption over the past few years has been focused on using Machine Learning (ML), Deep Learning and predictive models in combination with integrated data platforms containing official city records, logs, IoT sensor data and contextual information about physical assets. This has led to significant improvements in optimizing city operations through advanced analytics and predictive modeling [3]. Now, with GenAI, new capabilities can be introduced into these existing applications or new, transformative applications can be built by leveraging the ability of GenAI for pattern recognition and creation of novel content, ranging from NLG for human-like text to high-fidelity synthetic data for simulating planning scenarios [9].

This survey is structured as follows. In Section II, we establish the relevant foundational technologies used in a Smart City and the resulting data foundation built with these technologies. In Section III, we review how GenAI's fundamental capabilities can be broadly applied across different urban subsystems, and look at examples of specialized GenAI applications for performing specific tasks in individual urban subsystems. These areas include traffic management and simulation, energy management and grid simulation, urban planning, citizen services and public engagement. Additionally, we review examples of cities that have adopted GenAI applications based on integrated data platforms. We further examine how Foundational Models, by processing and integrating multimodal, multidimensional data present across the various urban subsystems can be used to build natural language based conversational interfaces that leverage the data foundation of city records, real-time IoT sensor data streams and Digital Twin virtual representations.

This is the first paper to the best of our knowledge that talks about GenAI applications in the context of the three main user archetypes in a Smart City - Citizens, Operators, and Planners. By surveying how conversational GenAI can bridge the gap between complex urban data infrastructure (IoT, Digital Twins) and diverse end-users, we aim to aid smart city development.

## II. FOUNDATIONAL TECHNOLOGIES FOR SMART CITIES

In this section, we explore the different technologies and data sources that underpin the creation of data-driven smart cities, and form the data foundation for using GenAI for urban applications.

### A. *Internet-of-Things and Digital Twins*

IoT is an umbrella term [3] that is used to define internet-connected sensor and control infrastructure for physical assets that enables data collection, monitoring and remote management of these assets [5] through integration with Supervisory Control and Data Acquisition (SCADA) systems. The value of creating an IoT-powered smart city lies in collecting real-time sensor data and using bidirectional commands to control and change the state of physical assets. Digital Twins (DTs) [11] are virtual counterparts of physical assets or systems. These digital representations continually adapt to operational changes based on real-time data collected about the physical processes associated with the asset [3] [6]. DTs are typically used to view, analyze and forecast the future of the assets through software applications [3].

### B. *Smart Cities using Urban Digital Twins*

A smart city can be conceptualized as a set of interconnected systems that continuously leverage new technologies to connect people, information, and urban infrastructure [4]. Urban Digital Twins *(*UDTs) in a Smart City are virtual representations of city systems connected through IoT technologies, with user interfaces for both citizen services, operational management and urban planning. These UDTs may be powered by real-time data and allow for autonomous bidirectional actions between the physical and digital entities [4]. UDTs integrate real-time data, historical records, and use predictive models to simulate, monitor, and optimize urban systems while offering a holistic view of the urban environment [6].

### C. *Data foundation of a Smart City*

Data generated in urban environments spans across UDTs in multiple subsystems and urban physical infrastructure [9]. This data is typically spatiotemporal and represents a constant state of flux due to ongoing human activities [11]. The Snap4City Smart City Digital Twin (SCDT) Framework [13] highlights the diversity of data generated by IoT devices in a smart city, and the challenges in integrating this data into a web based interface. Data generated in an urban environment has multiple modalities and dimensions, presenting various challenges in processing this data for feature extraction in the different urban subsystems [12].

### D. *Generative Artificial Intelligence (GenAI)*

GenAI refers to a set of AI techniques that uses models designed to learn the underlying patterns and structure of vast amounts of data and generate new data points that could plausibly be part of the original dataset. In the training phase, a generative model tries to estimate the probability distribution of data from which the training data is drawn. This learned understanding allows the model to subsequently generate new, synthetic data points that exhibit similar patterns to the original dataset [9].

The rapid development in the GenAI space brings a potential paradigm shift for managing and interacting with urban systems, with the possibility of extending the capabilities of existing applications and traditional AI/ML approaches. GenAI models are capable of consuming data from a single modality or multiple modalities such as text, images, video [5]. Leveraging the capability of GenAI to create digital artifacts across multiple modalities such as text, images and video allows for new, NLG based applications centered around conversational interaction for querying and summarizing datasets.

A few key technologies used in GenAI include Large Language Models (LLMs) [14] - based on Generative Pre-trained Transformers (GPT) [9], and synthetic data generation techniques such as Generative Adversarial Networks (GANs) [51], primarily used image generation, Variational Auto Encoders (VAEs) [50] and Diffusion Probabilistic Models (DPMs) [34]. In the following section, we will review the use of these techniques in building applications for each user archetype. The integration of multiple datasets to develop unified models is an active area of research, conceptualized under proposed frameworks like "Foundational Model" [16] [26] or "Large Flow Model" [17] for urban data, with promising results emerging from studies applying GenAI with UDT contexts [9][25].

## III. CONVERSATIONAL INTERFACES IN A SMART CITY

With human-like Natural Language Generation (NLG) capabilities and multilingual support, LLMs are a natural fit in solving various challenges faced in building smart cities. In this paper, we consider three critical user archetypes served through the creation of conversational interfaces using LLMs - **citizens, operators of urban subsystems, and urban planners**. Each of these users have a distinct set of expectations and requirements. By providing access to tools and applications that can provide holistic and situationally aware responses in human language, each of these users can discover information and interact with data more effectively.

### A. *User Archetype I: Citizens*

The Citizen archetype represents members of the general public who are typically seeking practical, often real-time or localized information (e.g., transit schedules, air quality, parking availability, local events) or needing to perform specific tasks (e.g., reporting issues, paying bills, finding services), navigate daily life, enhance convenience, ensure safety, or engage with their community. They require intuitive and accessible interfaces, like conversational AI, without needing deep technical understanding or control over the underlying urban systems. Applications based on conversational interfaces for citizens leverage both, NLG and NLU capabilities of LLMs. LLM usage by cities has shown promising results in providing citizen services, with high satisfaction rates for LLM responses in multiple languages [23] [36]. This subsection reviews some such examples.

Barcelona has adopted an integration platform at the municipal level with multiple LLMs trained on diverse source datasets spanning policy documents, transcriptions of parliament sessions, along with support for 4 languages [23], providing inclusive access of city services to all citizens.

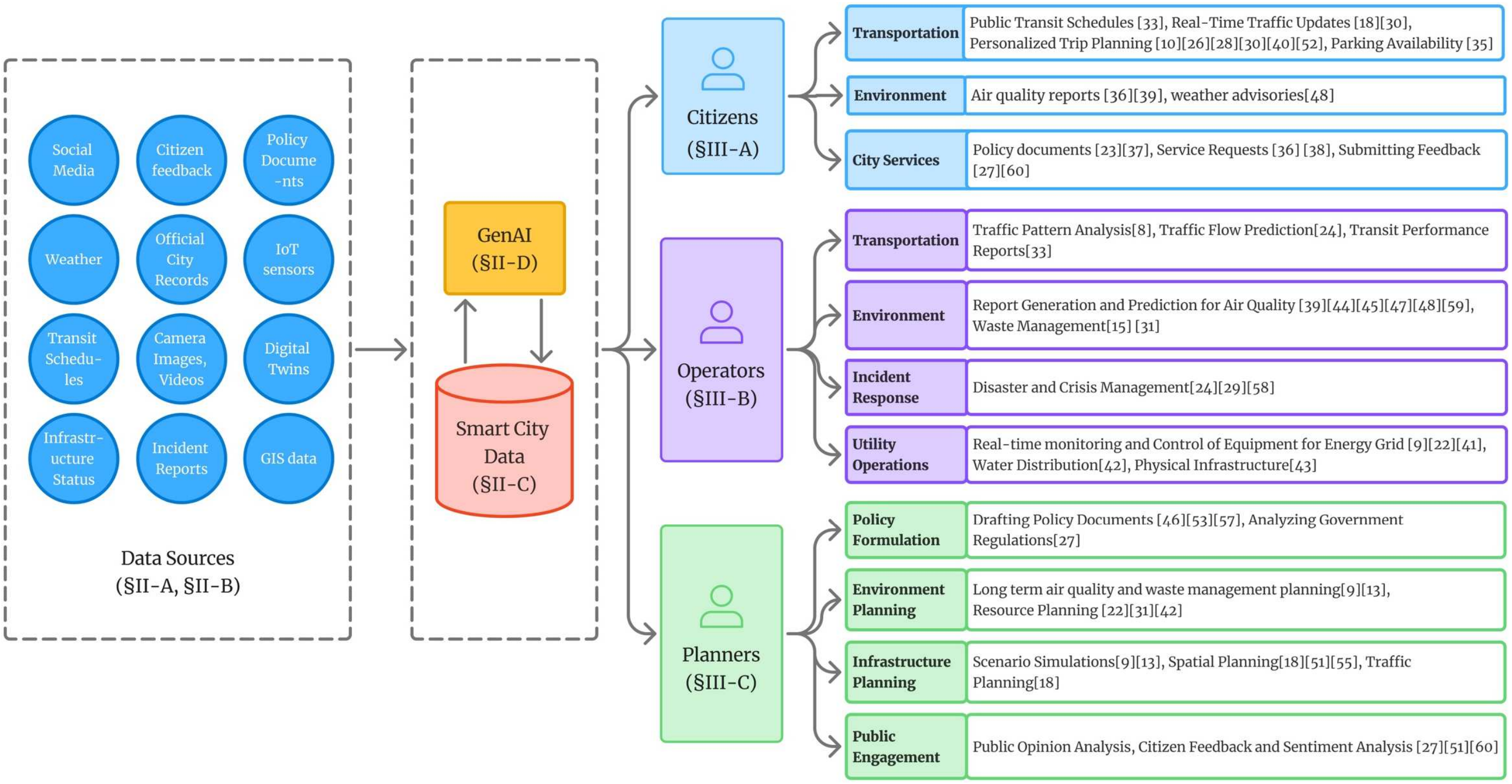

Figure 1: Opportunities and applications of Generative AI in integrating diverse smart city data to deliver tailored services for Citizens, Operators, and Planners.

In Copenhagen, LLMs have been deployed for processing public feedback and conducting sentiment analysis [27] [23]. Helsinki uses LLMs to provide information about urban services such as parking, sports, recreational facilities, rental housing, and also to analyze citizen feedback data to understand their needs. StadsbyggarenAI [60], a research project focused on the city of Malmö, enhances citizen engagement by translating complex planning documents and technical jargon into more accessible formats using an LLM, enabling a better understanding of ongoing and future urban developments, and a conversational interface to ask specific questions, and provide informed feedback. Barcelona has noted a 94% user satisfaction rate with responses generated with 30sec latency, while Vienna has reported a 42% increase [23] in first-time resolution of citizen service requests coupled with a 28% reduction [23] in service delivery costs. These initiatives are aimed at improving citizens' experience by making information more accessible, and increasing the level of citizen satisfaction while also enhancing the responsiveness of city services using GenAI.

In the areas of transportation and mobility, LLMs can help citizens in answering queries about public transit [33] schedules and providing real-time updates with natural language with grounding using Retrieval Augmented Generation (RAG) [33], which adds a retrieval mechanism to LLMs for retrieving data from external sources. A Foundational Model like the proposed UrbanLLM [26] can act as the central trip planning platform for citizens. Specific models such as CoDrivingLLM [10] can be used for decision making in driver assistance systems, while applications like DynamicRouteGPT [28] have been proposed for assisting citizens with route planning and simplification of complex transportation choices. Personalized systems like LLMob [52] can be used for generating mobility recommendations based on activity trajectories and movement patterns. Intelligent Transportation Systems can be built using multi-agent LLM-RAG powered models [40] for generating transport plans based on user-defined scenarios that reflect the personalized transportation and mobility needs of specific user groups, using both real-time [30] and historical data such as traffic conditions, weather, parking availability, and the user's current location.

For air quality and pollution monitoring, LLMs have been proposed to provide an integrated analysis and communication platform for citizens. An example in the area of air quality is VayuBuddy [39], an LLM-powered chatbot which provides users with answers and plots based on air quality data from IoT sensors from India's Central Pollution Control Board. By processing current air quality metrics, weather forecasts, and potentially even user-provided health information, LLMs can generate tailored advice and alerts. An example is the conceptual Instructor-Worker LLM system [48] which can provide specific health recommendations based on air quality analysis during an adverse event like wildfires, and provide citizens with information to take proactive steps like adjusting outdoor activities or using air purifiers based on personalized risk assessments derived from LLM analysis of current conditions.

While there are many opportunities to apply GenAI to improve the experience of citizens in Smart Cities, as we have seen above, there are also challenges associated with it, especially the risk of hallucination and incorrect response generation. Singapore's implementation of AI-Based Citizen Question-Answer Recommender (ACQAR) [36] has mitigated this through a Human-In-The-Loop (HITL) approach. ACQAR answers citizen inquiries with automated context-aware responses that consider the inquiry and predict emotional state, but a manual review is conducted by the city's Customer Service Officers (CSOs). The pilot study has shown a notable decrease in average case resolution time and improved citizen satisfaction rates in pilot studies.

Another mitigation technique used with LLMs to ensure factual responses is RAG [48] [49]. AskCOS, the conversational interface developed by The City of Colorado Springs [37] has a specialized knowledge base exclusively grounded in verified, city-controlled pages and documents sourced directly from the city's official website. The UK's GOV.UK Chat [38] is another example of an intelligent question-answering system designed to answer inquiries from both citizens and businesses, using an RAG grounded with the website's data.

Through building LLM-powered applications, a larger spectrum of citizens can interact with the urban environment and access services using human-like dialogue in a language they are fully comfortable with, thereby reducing barriers and improving equitable access to everyone.

### B. *User Archetype 2: Operators and Managers*

The Operators and Managers archetype includes professionals focused on the real-time monitoring, control and day-to-day management of specific urban subsystems like traffic, energy, or water, ensuring operational stability and responding to incidents. These users require access to detailed technical data, system alerts, and controls to maintain services and troubleshoot issues, differentiating them from citizens whose requirements are typically centered around seeking general information. Applications designed for operational management of urban subsystems leverage the advanced NLP capabilities of GenAI models for parsing and analyzing vast quantities of unstructured and structured data. LLMs can be used for tasks such as automated generation of summaries and synthesis of information into easy-to-understand text responses for operators and managers, across various situations. Unstructured data ranging from citizen feedback through surveys or social media, official reports, policy documents, and service request tickets [20][15]. As the IoT infrastructure in a city grows for sensing and control of physical assets, managing the vast amounts of structured data generated is a growing challenge [11] [19]. By breaking down traditional data silos [13] to combine IoT data with context from UDTs, LLMs can facilitate integrating and interacting with multi-modal, multi-dimensional spatiotemporal data from multiple urban subsystems. Examples include summaries of traffic incidents, traffic congestion patterns [8], transit performance [33], resource usage summaries for energy and water, environmental reports about air quality contextualized with weather forecasts [39], waste collection routes and smart bin levels [15] [31].

For quicker analysis and responses during times of crisis, LLM-based applications such as IncidentResponseGPT generate traffic incident response plans based on regional guidelines combined with real-time data [24], and DisasterResponseGPT [29] generates automated plans-of-action based on incident reports and allows for emergency operators to interactively explore a wide range of response options. Similar proposals suggest the use of decision assistance systems for natural disasters that use multimodal data across IoT, social media, and infrastructure status reports during events such as earthquakes and pandemics [56][58].

GenAI can power dynamic dashboards that proactively display the most pertinent information (critical parameters, events, mitigation options) to operators [22] based on real-time conditions, text queries, or alarms and adjust constantly to reflect the current operational reality. eGridGPT shows a real-life example of LLM usage in power grid operations [22], developed for aiding decision-making and managing the energy grid in an era with increasing complexity and uncertainty due to an evolving landscape of shifting loads and generation sources. This is a transformative shift away from traditional static displays which require operators to manually recall data locations across multiple monitors. Adoption of such interfaces can potentially streamline operator's focus on critical tasks by reducing cognitive load. Decision-support systems based on LLM-powered conversational querying interfaces can be used to assist operators of resources alongside UDTs [41]. For example, creating 3D models to represent UDTs that are regenerated to show current status using real-time IoT data [41]; or integrating real-time flow, pressure, and thermal imaging data to accurately predict water usage and detect network leakages [42]. These platforms can assist in analyzing vast amounts of textual data from regulations, operation manuals, maintenance logs and synthesizing IoT data from SCADA systems, to provide concise answers and potentially explain anomalies in the data in each urban subsystem by bridging the gap between unstructured technical information, spatiotemporal data, and human comprehension. This increases the situational awareness of operators and reduces time spent searching through complex data or documentation. Such GenAI platforms can be used to monitor performance, or query standard operating procedures, enabling operators to ask questions such as: "What's the water pressure trend in Zone 5 over the last 12 hours?" [43]. Summarization and explanations of anomalies can be similarly presented to operators in human-like language. Automated natural language reports can also be generated for buildings and physical infrastructure, based on data from sensors embedded in physical infrastructure to monitor their condition and usage patterns.

Forecasting and prediction is another area where LLMs can assist urban operations, evidenced by promising results seen in air quality monitoring and prediction models for different scenarios [44]. Examples include LLMAir [45] - which performs air quality prediction for PM2.5; Hybrid Enhanced Autoregressive Transformer (HEART) [47] - a proposal to predict air pollution; Instructor-Worker LLM - a multi-agent system designed for air quality prediction during wildfires [48]; Spatio-Temporal LLM (STLLM) - a solution for industrial PM2.5 forecasting using IoT sensors which relies on edge computing [59] to process data at the site instead of streaming to cloud systems [59]; UrbanGPT UrbanGPT [8] - a proposal to use LLMs with spatiotemporal data to predict traffic flow.

Challenges faced in building operator-facing applications include gaps in collected data, due to technical issues or equipment downtime. This can be mitigated by using data augmentation and imputation [9] techniques that enrich existing datasets to improve the training and robustness of downstream machine learning models to improve the accuracy of prediction models [50]. To ensure effectiveness of GenAI applications and present accurate data to operators for decision-making, RAGs can be used to ground models with combined datasets from IoT data streams and UDT definitions [49] [42].

### C. *User Archetype 3: Urban Planners*

The Urban Planners archetype includes users responsible for the city's long-term strategic development and planning. These users are typically focused on future growth, sustainability, land use, spatial design, and policy formulation. Requirements center around utilizing aggregated data, observe trends over time, analyze policy documents, and use simulation tools. Common tasks include analyzing complex urban dynamics, evaluating potential scenarios, and shaping the city's overall trajectory, distinct from operators managing real-time systems or citizens seeking general information. The needs of users tasked with urban planning range from information discovery, drafting documents, collaborative planning and scenario simulations and planning. CityGPT [25] is an example of an assistance tool for urban planners which allows querying urban knowledge graphs and assists in understanding spatio-temporal patterns of human mobility in the city context. LLMs enhanced with RAG can ingest specific local documents [46] like smart city master plans and government regulations to provide decision support for urban development and generating planning reports and draft policy documents [57] [53].

A proposal to create an Urban Generative Intelligence (UGI) [58] integrates LLMs with a city simulator and Urban Knowledge Graph (UrbanKG) to create a foundational system for urban intelligence. Using embodied agents within the simulation, realistic modeling of complex urban dynamics like mobility can be performed through a natural language interface.

Another key capability of GenAI models is generating synthetic data that mimics real-world conditions to build simulations of scenarios that might occur in various urban subsystems. Urban planners can leverage such synthetic data to test the potential impacts of different development plans and policies in a simulated environment before real-world implementation. This allows for exploring simulated "what-if" scenarios, one such instance could be “what if a city introduces a new policy to convert 20% of its downtown on-street parking spaces into dedicated bicycle lanes over the next five years” [9] [13]. Such simulations could help understand emergent system behaviors resulting from planning decisions. An example is planning for changes in traffic management approaches using AI-generated data for different simulations of traffic scenarios like congestion levels, incident occurrences and driver behavior changes. This data can help in enhancing the accuracy of traffic flow prediction, enabling proactive traffic management strategies [18]. Another example is the use of GenAI to generate synthetic air quality data to fill in gaps in air quality sensor networks [9] or to create more diverse training datasets for prediction models. Simulations can be created to evaluate the impact of different strategies such as traffic restrictions, emission controls, enabling proactive interventions to mitigate pollution.

Traffic scenario simulations that include traffic flow, rare events such as accidents and driver behavior modeling [18] can be used for informing decisions for improving traffic conditions [31]. By training systems using simulated data, mitigation strategies for adverse conditions can be tested in a risk-free UDT environment before actual deployment. Similarly, to improve management of energy infrastructure, buildings and other physical infrastructure, predictive maintenance models can be trained with simulated data for load patterns [9].

For collaborative participatory planning, PlaceMakingAI [51] and Land Use Configuration GAN [31] are examples of how GANs can be used to generate realistic, synthetic visualizations of potential urban spaces with real-time images for providing an interactive interface for citizens, designers, and stakeholders to collaboratively envision street design improvements and facilitate better decision-making for spatial planning [55]. Planners can perform automated sentiment analysis with LLMs to gauge public opinion on infrastructure projects or policies and make planning decisions based on emerging citizen concerns gathered through citizen feedback surveys, social media, and service request tickets [27].

## IV. Discussion

### A. Challenges

Despite promising advantages, multiple challenges exist in effectively implementing applications using GenAI. These challenges include data collection, storage, processing and management for an IoT enabled city due to the sheer volume and diversity [19] of data. Utilizing big data management techniques [19], along with an integrated IoT and UDT data platform 13] and offloading data processing to edge computing [59] can help to alleviate these issues. Gaps in collected data due to technical issues and a lack of representative data for rare events are another challenge. GenAI techniques such as VAEs, GANs and DPMs can be used for imputation and augmentation of such missing data [9].

GenAI models face limitations in their capabilities to perform numerical interpretation and may generate incorrect responses known as hallucinations. To accomplish successful adoption of these conversational interfaces, it is essential to mitigate current model limitations, as seen in the case of New York City’s MyCity LLM chatbot, which faced hallucination issues that caused the generation of incorrect and illegal responses during beta testing in 2023 [54]. To significantly enhance model trust and flexibility, hallucinations can be minimized through factual grounding of the LLMs through RAG [49], introduction of rule based constraints, and enabling continuous learning to keep models up-to-date on urban dynamism. Care must be taken to avoid the introduction or amplification of biases, and maintaining privacy of citizens when training GenAI models. The computational overhead and resulting costs associated with training and performing inference with GenAI models may be a roadblock due to limited city budgets and further limit the ability to integrate GenAI at scale [21]. Difficulties also exist in ensuring the fidelity of the synthetic data to accurately reflect real-world distributions [9]. Validation steps must be introduced to verify synthetic data.

Beyond technical issues listed above, we also face societal adoption and acceptance challenges due to the evolving nature of Human-AI collaboration techniques [32] [31]. Careful evaluation and benchmarking of GenAI models is necessary to ensure the usability and effectiveness of these technologies in real-world applications [20]. Establishing transparent policies and ethical guidelines while adopting AI is a key step towards establishing trust in the system [23]. A particular focus must be placed on ensuring human-in-the-loop [52] [58] decision workflows that allow for human oversight and validation of AI decisions.

## B. *Future Direction*

Deploying a robust GenAI framework will transform UDTs into truly dynamic frameworks for monitoring, improving and enhancing sustainable urban development [9]. To build upon the current capabilities of GenAI and address the challenges discussed above, future research should be focused on developing unified GenAI platforms designed to work across the complexities of the urban environment, unlocking the potential to build a central, easy-to-use platform that abstracts the inherent complexity and multimodality of spatiotemporal data in interconnected urban systems [1]. Such a platform should be capable of seamlessly integrating diverse IoT datasets into a text based platform while capturing trends and dependencies over a longer time horizon, such as the related effects of traffic congestion on air quality [16]. Beyond improved integration and analysis, another critical step is advancing GenAI towards reliable predictive generation for simulating and assessing potential bottlenecks of plausible future urban scenarios [13], and potentially integrating these simulation tools into interactive platforms [51] such as 3D virtual city models [9][41].

# V. Conclusion

GenAI offers transformative potential in reshaping data interaction within Smart Cities. By leveraging rich data from IoT and Urban Digital Twins, GenAI can offer intuitive interfaces to enhance citizen access to services, improve operational efficiency, and support strategic urban planning. In this paper, we present a comprehensive user centric survey of GenAI applications within the context of three critical user archetypes in a Smart City: Citizens, Operators/Managers, and Planners. We discuss different foundational technologies for smart cities. Further, we detail out how GenAI enables conversational interfaces tailored to the specific needs and applications for the three key user archetypes Lastly, we discuss the key challenges in implementing GenAI for smart cities and outline potential future directions for research and development. By surveying how GenAI can leverage existing IoT and Digital Twin data foundations, this paper aims to aid smart city development by making complex urban data more accessible and actionable for diverse stakeholders.